\documentclass[twocolumn,showpacs,preprintnumbers,amsmath,amssymb]{revtex4}

\usepackage{graphicx}
\usepackage{dcolumn}
\usepackage{bm}
\usepackage{color}

\begin{document}

\preprint{PRE}

\title{Universal Scaling of Optimal Current Distribution in Transportation Networks}

\author{Filippo Simini}
 \email{simini@pd.infn.it}
\author{Andrea Rinaldo$^2$}
 \email{andrea.rinaldo@epfl.ch}
\author{Amos Maritan}%
 \email{maritan@pd.infn.it}

\affiliation{%
Universit\`{a} di Padova, Dipartimento di Fisica ``Galileo Galilei"\\
via Marzolo 8, CNISM and INFN, 35131 Padova, Italy.\\
$^2$ECHO/ISTE/ENAC, Ecole Polytechnique F\'ed\'erale, 1015 Lausanne, Switzerland 
}%

\date{\today}

\begin{abstract}
Transportation networks are inevitably selected with
reference to their global cost which depends on the strengths and
the distribution of the embedded currents. We prove that
optimal current distributions for a uniformly injected
$d$-dimensional network exhibit robust scale-invariance properties,
independently of the particular cost function considered, as
long as it is convex. We find that, in the limit of large currents,
the distribution decays as a power law with an exponent equal to $(2d-1)/(d-1)$. The current distribution can be exactly
calculated in $d=2$ for all values of the current. Numerical simulations further suggest
that the scaling properties remain unchanged for both
random injections and by randomizing
the convex cost functions.
\end{abstract}

\maketitle

\section{Introduction}

Finding efficient ways of distributing (or collecting)
matter injected through a given region, spanned, e.g., by a regular lattice, from
(at) a unique source (sink) is
relevant to a variety of problems arising both for natural and
artificial systems. The main mechanisms that are known to achieve a capillary
distribution are by diffusion or through a network-like structure providing a near uniform spatial
supply (drainage), or a combination of them.
Network arrangements, of this kind, are observed in many living
organisms, like for instance circulatory and lymphatic systems in animals or
xylem and roots in vascular plants, and are also widely
employed in artificial systems such as electrical or hydraulic
transmissions and fluvial basins \cite{mcmahon,mays,ball,rinaldoiturbe,caldarelli}.

Certainly, one
wonders what is the basic selection principle that favors, say,
tree-like versus looping network structures, in view of the
widespread occurrence of both forms in nature and elsewhere \cite{barabalb,barabl}
\cite{caldarelli}. To that end, it has been previously shown that
tree-like structures emerge as local minima of global energy
expenditure in networks whose transportation cost is physically
constrained to be a concave function, such as in the  case of river
networks \cite{banavar2001}.

Transportation costs generally depend on the strength of currents and on network topology.
The best known example is the electrical resistor network \cite{doyle, straley}: consider a square lattice where a
resistor is placed at every bond between each pair of
nearest-neighbor nodes. Each node is externally supplied by a unit
flux. If a sink collects all the currents, one is
capable of controlling the current fluxes in all bonds by
assigning the potential differences between pairs of nodes. The total cost in the
transportation -- the dissipated power -- is proportional by Ohm's law to the sum
of the square of currents: $\mathcal{E}\left(\{I_b\}\right)=\sum_{b} |I_b|^2$. Optimal resistor networks are thus obtained by setting the potentials in
order to minimize the total transportation cost. The most
immediate way to generalize the resistor network case is
to replace the exponent ``$2$" by a generic exponent $\gamma > 0$.\\
The case $\gamma \leq 1$ has been characterized exactly
\cite{rinaldoiturbe,banavar2001, maritan2, maritan, maritan2000}.
It was shown there that the cost function admits many local minima
corresponding to configurations with currents present only on the
bonds of spanning trees \cite{banavar2001}. The case $\gamma=1/2$
exhibits scaling behavior akin to river networks
\cite{IRI,maritan,maritan2,rinaldoiturbe}.\\
The case $\gamma=1$ is
related to the Voter model \cite{liggett}, mass aggregation
\cite{swift, taka88,taka91,taka94}, directed sandpile models
\cite{dhar}, and Kleiber's law of metabolic scaling of living
organism \cite{nature}.
Recently optimal transportation networks with a global constraint have been studied in a variety of contexts \cite{durand,magnasco}, where different topologies of the optimal transportation network arising in the cases $\gamma<1$ and $\gamma>1$ have been investigated also by Bohm and Magnasco \cite{magnasco}.\\
The case $\gamma>1$ is an example of convex transportation costs. Important real-world examples are found, for
example, in road traffic analyses where more cars cause disproportionately higher costs (travel times), usually modeled as convex functions or in any electricity distribution
network owing to Ohm's law. Convex cost functions also occur in many operations research applications as pointed out in \cite{Golden}.

\section{Numerical Results}

\begin{figure}
\begin{center}
\includegraphics[height=55mm]{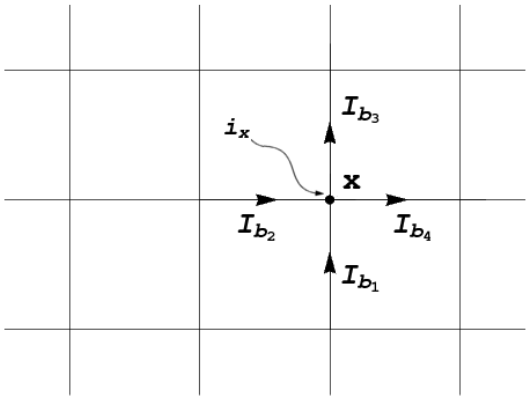}
\caption{The current conservation at node $x$. The sum of all currents flowing into node $x$ must be equal to the sum of currents flowing out of node $x$: $I_{b_1}+ I_{b_2}-I_{b_3}-I_{b_4}+i_x=0$.} \label{lattice}
\end{center}
\end{figure}

Here we address the current distribution for the case
corresponding to a general convex cost function of the circulating currents:
\begin{equation}\label{001}
\mathcal{E}\left(\{I_b\}\right)=\sum_{b} E(\mid I_b\mid)
\end{equation}
where the sum spans all network bonds $b$.
In specific examples and numerical simulations we will consider the particular class of convex functionals with
$E(\mid I_b\mid)=\mid I_b\mid^{\gamma}/\gamma$ and $\gamma>1$. It
will be shown, however, that our findings are valid for an arbitrary convex cost function $E$ with finite first derivative which depends only on current strength. Notice that for a non-linear resistor network the transportation cost $\mathcal{E}$ is proportional to the dissipated power only for the case $E(\mid I_b\mid)\propto\mid I_b\mid^{\gamma}$, as discussed in \cite{michele,millar}. We will focus here only on finite networks.
The goal is to determine the current probability distribution corresponding to the current
configuration $\{I_b\}$ which minimizes the total transportation cost $\mathcal{E}$
in the large size limit and its scaling behavior as a function of the cost $E(\mid I\mid)$.\\
The minimization of $\mathcal{E}$ is subject to the local
constraints of current conservation at each node $x$: the sum of
all currents flowing into a node, taken as
negative (positive) if directed outward (inward), must equal
the injected nodal flux, $i_x$ (see Fig. \ref{lattice}).

\begin{figure}
\begin{center}
\includegraphics[height=65mm]{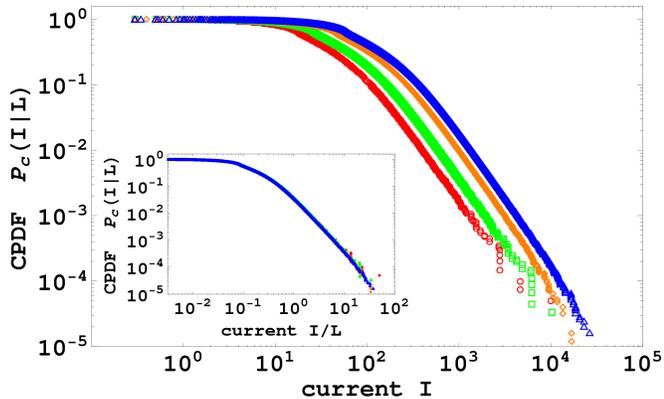}
\caption{(Color online) CPDF for square lattices of different sizes
($L=201$ (red circles), $301$ (green squares), $501$ (orange rhombi), $701$ (blue triangles)) and for $\gamma=2$ (on log-log scale).
Fitting the tail of the distribution with a power law yields a
value for the exponent $\tau\simeq 3$, e.g. for $L=301$ we get
$\tau=2.975\pm 0.045$ (the exponent of the power law was estimated
using the method of maximum likelihood \cite{likeli,newman}, and
the error was calculated with the bootstrap method \cite{boot}).
In the inset we plot $P(I|L)$ vs $I/L$, for all $L$'s and $I$'s
considered above. The collapse of the curves indicates that the
CPDF is a function of the ratio $I/L$.} \label{cumul2}
\end{center}
\end{figure}

Because such constraint is linear in the $I_b$s,
if $\mathcal{E}$ is a  convex function
then it admits a single global minimum.
The existence and uniqueness of
the solution for infinite networks have been addressed elsewhere \cite{michele}.\\
As shown below, we find quite generally that the cumulative probability distribution function (CPDF),
i.e., the fraction of currents $|I_b|$ larger than $I$, obeys
the finite--size scaling
\begin{equation}\label{02}
P_c(I|L)= F\left(\frac{I}{L}\right)\quad \forall d, \gamma>1,\ I\gtrsim 0
\end{equation}
where $L$ is the linear size of the system and $F$ is the scaling function which at large $x$ behaves as $F(x)\sim
x^{1-\tau}$ with $\tau=(2d-1)/(d-1)$ and
$\lim_{x\rightarrow 0}F(x)=1$. This scaling form holds independently of the particular convex cost function considered. The standard scaling behavior for the case
$E(\mid I_b\mid)=\mid I_b\mid^{\gamma}/\gamma$ and $\gamma\leq 1$ \cite{maritan2,colaiori1997} corresponding to a non-convex cost function, was instead found to be:
\begin{equation}\label{011}
P_c(I|L)=I^{1-\tau}f\left(\frac{I}{L^d}\right)\quad \gamma<1,\ I\gtrsim 1
\end{equation}
with $\lim_{x\rightarrow 0}f(x)=$ const, i.e., a pure power law is
obtained in the large size limit ( $\tau = 1.43\pm 0.03$ in $d=2$
and $\gamma=1/2$).
The $\gamma=1$ case was solved exactly in all
dimensions using a mapping to reaction diffusion models
\cite{swift} and  $\tau=2(d+1)/(d+2)$ when the dimensionality is
lower than the upper critical dimension, $d_c = 2$, and $\tau=3/2$
when $d>2$. In the present case no upper critical
dimension is found above which the exponent remains the same.\\
Let us first describe the results of the numerical determination of
current configuration minimizing Eq.(\ref{001}) with $E(\mid
I_b\mid)=\mid I_b\mid^{\gamma}$ and $\gamma >1$ in a square
lattice (i.e., $d=2$) of linear size $L$ where a uniform input at each
site $i_x=1$ is assumed (at the sink where
all currents are collected one has $i_{sink}=-L^2$+1). We have used open boundary
conditions for simplicity because we do not expect that they
influence the scaling behavior in the large size limit.
Because our problem reduces to the minimization of a
convex function of many variables, we have used the nonlinear
conjugate gradient method \cite{conjgrad}. The CPDF, $P_c(I|L)$, is plotted in Fig. \ref{cumul2}
for lattices of different sizes $L$ and for $\gamma=2$, the
resistor network. It has the following scaling behavior:
\begin{equation}\label{01}
P_c(I|L)=\begin{cases}
\textrm{const} & \textrm{for}\: 0\lesssim I \lesssim L\\
I^{1-\tau} & \textrm{for}\: L\lesssim I \lesssim L^2\end{cases}
\end{equation}
with $\tau=2.975\pm 0.045$. In the inset of Fig. \ref{cumul2}
the CPDF is plotted versus $I/L$ for various $L$, showing that
indeed $P_c(I|L)$ is a homogeneous function of the ratio
$I/L$.

\begin{figure}
\begin{center}
\includegraphics[height=90mm]{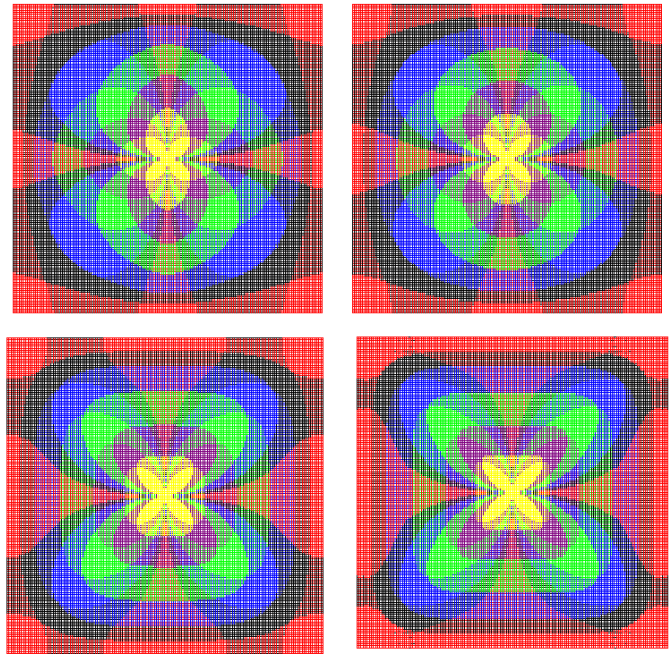}
\caption{(Color online) Currents' intensity in the optimal configuration for
different values of the exponent $\gamma$ and for the size $L=151$ of a square lattice.
From top left to bottom right: $\gamma=1.5$, $\gamma=2$,
$\gamma=4$ and $\gamma=6$. The petal-like arrangement of currents is not an artifact of the underlying lattice geometry but arises from the decomposition of current vectors into components.
The direction of currents is towards the center (directed networks). The colors indicate the
intensity of currents: Yellow: $L \leq I$, Purple: $L/2 \leq I <
L$, Green: $L/4 \leq I < L/2$, Blue: $L/8 \leq I < L/4$, Black:
$L/16 \leq I < L/8$, Red: $I < L/16$.} \label{intensita}
\end{center}
\end{figure}

\begin{figure}
\begin{center}
\includegraphics[height=65mm]{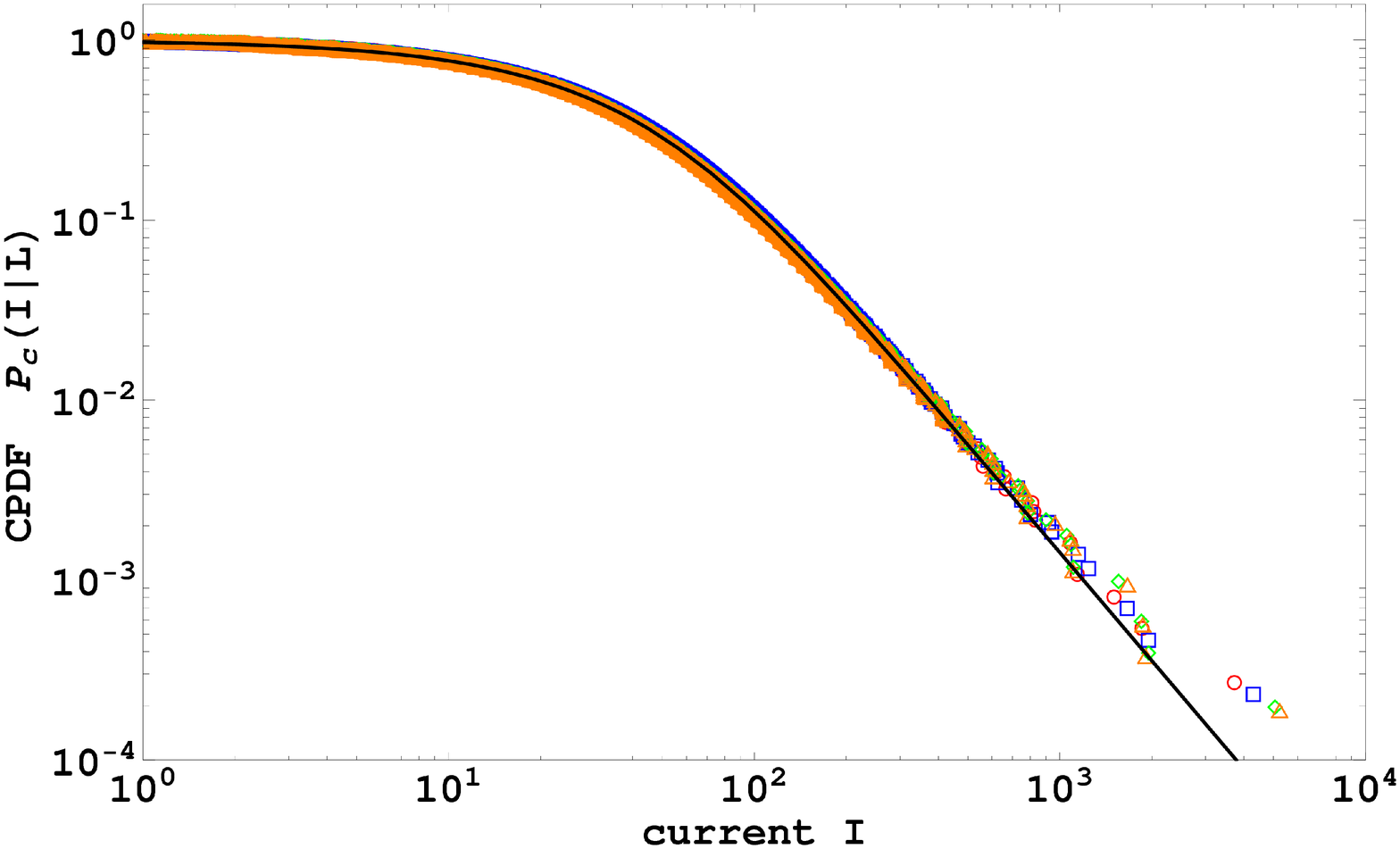}
\caption{(Color online) Comparison of CPDFs of global minimum configurations for the class of convex functionals (\ref{001}) with $E(\mid I_b\mid)=\mid I_b\mid^{\gamma}/\gamma$ and different $\gamma$-values:
$\gamma=1.5$ (red circles), $2$ (blue squares), $4$ (green rhombi), $6$ (orange triangles). The black solid line is eq.(\ref{10}), the
CPDF of the analytic solution of eq.(\ref{6}). The shapes of the network boundaries in the various cases are chosen as explained in the text.} \label{varigamma}
\end{center}
\end{figure}

We have also performed numerical optimizations with different values of $\gamma>1$ in $d=2$. Figure \ref{intensita} shows pictures of the optimized current configuration and the corresponding CPDFs are plotted in Fig. \ref{varigamma}. Because data overlap, and although the
current configuration of global minimum for $\mathcal{E}$ varies
with $\gamma$ as shown in Fig. \ref{intensita}, it is suggested
that the distribution of currents is independent of the exponent $\gamma$ when $\gamma>1$.
Additionally, we performed simulations on a $2-d$ triangular lattice. The scaling behavior of the CPDF proved to be independent of the underlying lattice's structure.\\

\section{Analytical results}
We attack the problem analytically in the continuum limit
(this will be justified a posteriori). We begin to illustrate
the procedure in detail for the case $E(\mid I_b\mid)=\mid I_b\mid^{\gamma}/\gamma$; later on we will extend it to the more general case.\\
Under the assumption that the current distribution, in the large size limit, does not depend on the shape of the volume we enclose our system in a region $\Omega=\{x:\
\|\mathbf{x}\|_{\gamma}\leq L \}$, whose volume will be
denoted as $|\Omega|$. We have defined the norm as
$\|\mathbf{x}\|_{\gamma}\equiv[\sum_{\mu}|x_{\mu}|^{\gamma}]^{1/\gamma}$. In
the case $\gamma=2$, $\Omega$ is a sphere of radius $R\equiv
L$. Let $\mathbf{j}(\mathbf{x})$ be the current
density at location $\mathbf{x}$ whereas
$i(\mathbf{x})=i_0\left[1-|\Omega|\delta^d\left(\mathbf{x}\right)\right]$
is the external input. The Dirac delta distribution $\delta^d\left(\mathbf{x}\right)$ [a compact notation for the $d$-dimensional notation $\delta^d\left(\mathbf{x}\right)\equiv \delta(x_1)\ldots \delta(x_d)$ ] represents the sink at the origin, whereas $i_0$ is the uniform input. The components
$j_{\mu}(\mathbf{x})\quad (\mu=1,...,d)$ of the vector
$\mathbf{j}(\mathbf{x})$ represent the currents along the positive
direction of coordinate axes at position $\mathbf{x}$.

In the continuum case we define a cost functional analogous to Eq. (\ref{001}) as:
\begin{equation}\label{5}
    \mathcal{E}=\int_\Omega \textrm{d}^dx\ \frac{\|\mathbf{j}(\mathbf{x})\|_\gamma^{\gamma}}{\gamma}.
\end{equation}

We search for the current configuration that minimizes the cost function Eq.(\ref{5})
with the constraint of the current conservation law, $\nabla \cdot
\mathbf{j}(\mathbf{x})=i(\mathbf{x})$, at each position, $\mathbf{x}$. This is done by introducing a Lagrange multiplier (potential), $V(\mathbf{x})$,
at each position, $\mathbf{x}$, and solving the following equation:
\begin{equation}\label{6}
   0= \frac{\delta}{\delta j_\mu(\mathbf{x})}\left(\mathcal{E}+\int_\Omega \textrm{d}^dx\ V\nabla \cdot  \mathbf{j}\right)= \frac{j_\mu(\mathbf{x})}{|j_\mu(\mathbf{x})|^{2-\gamma}}-\frac{\partial}{\partial x_\mu}V(\mathbf{x})
\end{equation}
where $\mu=1,2,\dots, d$. Because we expect that the CPDF
does not depend on boundary conditions in the large size limit, we
choose $\mathbf{j}(\mathbf{x})=0$ at the boundary. We now assume that the solution depends only on
$\|\mathbf{x}\|_{\gamma}$ : because $\mathcal{E}$ is convex, a solution with this property (if it exists) is the unique solution. Using the above choice of the input currents $\int_\Omega \textrm{d}^dx \; i(\mathbf{x}) =0$ and the conservation law $\nabla \cdot
\mathbf{j}(\mathbf{x})=i(\mathbf{x})$, by applying Gauss theorem one gets that the boundary condition on the volume $\Omega$ are automatically satisfied. Equation (\ref{6}) together with current conservation gives the following radial current density as the optimal solution of
Eq.(\ref{5})
\begin{equation}\label{8}
    \mathbf{j}(\mathbf{x})=\mathbf{x}\frac{i_0}{d}\left(1-\left(\frac{L}{\|\mathbf{x}\|_{\gamma}}\right)^d \right).
\end{equation}

This solution is radially symmetric only with the metric defined in terms of the $\gamma$-norm itself: thus the equi--currents lines defined by $\|\mathbf{j}\|_{\gamma}=$ constant are circles only for $\gamma=2$, whereas when $\gamma\to\infty\ (1)$ they become squares with sides parallel to the coordinate axis ($45^{\circ}$--tilted squares).
The CPDF is given by $P_c(j|L)=|\Omega|^{-1}\int_\Omega \textrm{d}^dx\ \sum_{\mu}\Theta\left(|j_\mu(\mathbf{x})|-j\right)$ if we consider currents' components, or by
\begin{equation}\label{9}
    P_c(j|L)=|\Omega|^{-1}\int_\Omega \textrm{d}^dx\ \Theta\left(\|\mathbf{j}(\mathbf{x})\|_{\gamma}-j\right)
\end{equation}
if we consider current norm \footnote{It can be easily shown that if we define the CPDF with reference to a generic norm $\|\cdot \|_{\alpha}$ the asymptotic behavior at large currents remains unchanged.} [$\Theta(z)=1$ if $z>0$ and zero
otherwise]. It can be shown that asymptotic behaviors are the same in both cases.
Using the explicit solution Eq.(\ref{8}) one sees that  $P_c(j|L)$ depends only on the dimensionless ratio $j/(i_0L)$ for all $d$. When $d=2$ Eq.(\ref{9}) takes the simple closed form:
\begin{equation}\label{10}
   P_c(j|L)=F\left(\frac{j}{i_0L}\right), \quad F(z)=\left( \sqrt{1+z^2}-z\right)^2
\end{equation}
which is of the kind anticipated in Eqs.(\ref{01}) and (\ref{02}) with $\tau=3$. The prediction Eq.(\ref{10}) is shown in Fig. \ref{varigamma} compared to CPDFs of numerical simulations calculated considering currents' components. Even though we do not have the explicit analytical form for $d>2$ it is not difficult to verify the asymptotic behavior of Eq.(\ref{01}). Indeed for $j/L\gg 1$ the leading contribution in Eq.(\ref{9}) comes from $\|\mathbf{x}/L\|_{\gamma}\ll 1$ leading to $P_c(j|L)\sim (j/L)^{(1-\tau)}$ with $\tau= (2d-1)/(d-1)$ for $d>1$. The minimum $\|\mathbf{x}\|_{\gamma}$ is given by the underlying lattice spacing, say $a$, and so, according to Eq.(\ref{8}), the maximum current is of order $ai_0(L/a)^d$. Thus scaling holds in the region $1 \lesssim j/i_0L \lesssim (L/a)^{d-1}$. The special case $d=1$ is trivial and one gets $P_c(j|L)=1-2j/(i_0L)\Theta(1-2j/(i_0L))$. Notice that in this case the scaling region has shrunk to zero.\\
The case $\gamma<1$ cannot be treated in the same way as above. In fact in \cite{maritan2000} it was shown that for the functional (\ref{001}), with $E(z)\propto z^{\gamma}$, any current configuration, with currents being different from zero only on the bonds of a spanning tree, is a local minimum. For such solutions the second term in Eq.(\ref{6}) would diverge in correspondence of the bonds not belonging to the spanning tree.\\
The above results can be generalized to the case of a generic convex cost function with finite first derivative as follows.

While in the presence
of an underlying network the natural choice for the cost function is
given by eq. (\ref{001}) in the continuum there are at least two
natural choices. The first choice corresponds to
$\mathcal{E}=\int_\Omega \textrm{d}^dx\ \sum_{\mu}
E(|j_{\mu}(\mathbf{x})|)$ whereas the second one is given by:
\begin{equation}\label{5b}
    \mathcal{E}=\int_\Omega \textrm{d}^dx\ E(\|\mathbf{j}(\mathbf{x})\|_{\gamma}).
\end{equation}
It can be shown that, although these two functionals have
different optimal configurations, their CPDF have the same scaling
behavior for small and large currents. For the previous case, $E(z)= z^{\gamma}/\gamma$,
the two choices coincide.
The minimum of the cost function, Eq.(\ref{5b}), in the domain
$\Omega=\{x:\ \|\mathbf{x}\|_{\gamma}\leq L \}$ and with the constraint of current conservation proceeds as before.
The stationarity conditions of the constrained problem are:
\begin{equation}\label{general}
E'\left(\|\mathbf{j}(\mathbf{x})\|_{\gamma}\right)\frac{|j_\mu(\mathbf{x})|^{\gamma-2}}
{\|\mathbf{j}(\mathbf{x})\|^{\gamma-1}_{\gamma}}j_\mu(\mathbf{x})=
V'\left(\|\mathbf{x}\|_{\gamma}\right)\frac{|x_\mu|^{\gamma-2}}{\|\mathbf{x}\|^{\gamma-1}_{\gamma}}x_\mu 
\end{equation}
where, as before, we have assumed that the potential $V(\mathbf{x})$ is a function of $\|\mathbf{x}\|_{\gamma}$. The solution is given by $j_{\mu}(\mathbf{x})=x_{\mu}f(\|\mathbf{x}\|_{\gamma})$ with $V(z)$ satisfying the equation $E'(f)=V'(z)$ (\textit{prime} indicates the derivative with respect to the argument). Imposing current conservation we get $f(z)=i_0/d(1-(L/z)^d)$ leading again to solution (\ref{8}).
In turn this implies that the scaling behavior of the CPDF as defined in Eq.(\ref{9}) is independent of the specific cost function as long as it remains convex.\\

\section{Inhomogeneous cases}

\begin{figure}
\begin{center}
\includegraphics[height=55mm]{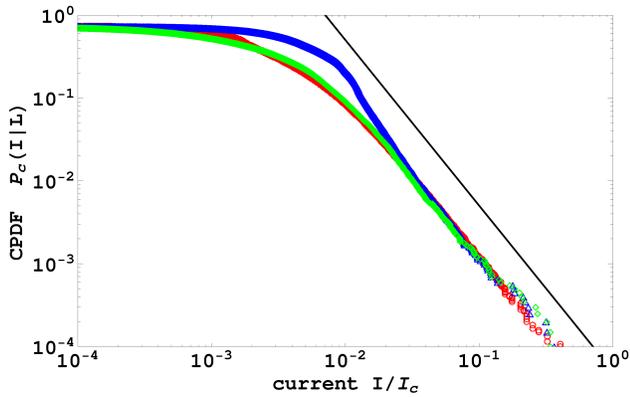}
\caption{(Color online) Comparison among the CPDF for the uniform case (red circles) with two examples of the heterogeneous conductivity (green rhombi) and injection cases (blue triangles) for a $L=151$ lattice, with $E(\mid I_b\mid)=\mid I_b\mid^{\gamma}/\gamma$ and $\gamma=2$. The probability distribution used to extract random resistances and injections is a power law with exponent $-1.5$, at large values, in order to provide a high degree of inhomogeneity. $I_c$ is properly chosen to show that the power law exponent is the same at large currents for the three configurations. The shown straight line has slope $-2$ as our analytical results predict.}
\label{rand}
\end{center}
\end{figure}

We have further tested the robustness of our results by
performing additional numerical simulations on systems subject
to independent, equally distributed random current injection,
$i(\mathbf{x})>0$ at the nodes, or in the presence of non--uniform conductivity where the cost function is given by
$\mathcal{E}(\{I_b\}) \ = \ \sum_b k_b |I_b|^{\gamma}$
where $k_b$ are random positive numbers.\\
As random distribution for injections and conductances we have
chosen a power law to ensure a high degree of
inhomogeneity. The simulation results of Fig.
\ref{rand} show that the leading trend for large
currents remains the same as in the uniform case studied
above. Thus it is plausible that the scaling behavior of the
current distribution corresponding to the optimal solution of the
uniform case might remain the same even for the more general case
of a spatially varying convex cost function.
These results differ from the case of
random transportation dynamics \cite{taka88,taka91,taka94} for which it
was shown that the uniform injection case is equivalent to our optimization problem with $\gamma=1$ \cite{maritan2}.
Indeed for these models the scaling behavior of the CPDF proves sensitive
to the distribution of the injections. However the present numerical results suggest
that this equivalence can not be generalized to the random injection case.\\

\section{Conclusions}

In summary, we have studied a class of optimal transportation
networks with a convex cost function as given by Eq.(\ref{001}) whose
prototype is
$\mathcal{E}\left(\{I_b\}\right)=\sum_{b} \mid I_b\mid ^{\gamma}$
with $\gamma>1$. The optimal current configurations exhibit a
probability distribution function characterized by a scaling behavior
given by Eqs.(\ref{02}) and (\ref{01}). The scaling
exponent of the current distribution proves robust with
respect to: (\textit{i}) the choice of the transportation cost, as far
as it is convex and has finite
first derivatives with respect to the currents; (\textit{ii}) the
distribution of injected currents; (\textit{iii}) position--dependent
(convex) cost functions. The analytical
results show that the exponent of the asymptotic power--law behavior of
the current probability distribution function varies
continuously from $3$ in two dimensions to $2$ at infinite dimensions
with no evidence of an upper critical dimension.\\

\noindent {\bf Acknowledgements} We are grateful to
Jayanth Banavar for invaluable discussions.
This work was
supported by a grant of Fondazione Cassa di Risparmio 2008.


\end{document}